\documentclass{article}
\usepackage[vmargin=3cm,hmargin=3cm]{geometry}
\usepackage{amsmath}
\usepackage{amssymb}
\usepackage{authblk}
\usepackage{rotating}
\usepackage{url}

\author[1]{Rakhi Manohar Mepparambath}
\author[1,*]{Hoai Nguyen Huynh}

\affil[1]{Institute of High Performance Computing (IHPC),

Agency for Science, Technology and Research (A*STAR),

1 Fusionopolis Way, \#16-16 Connexis, Singapore 138632,

Republic of Singapore}

\affil[*]{Corresponding author

Email: huynhhn@a-star.edu.sg}

\date{}

\begin{document}

\title{The role of spatial scales in assessing urban mobility models}

\maketitle

\begin{abstract}
Urban mobility models are essential tools for understanding and forecasting how people and goods move within cities, which is vital for transportation planning. The spatial scale at which urban mobility is analysed is a crucial determinant of the insights gained from any model as it can affect models' performance. It is, therefore, important that urban mobility models should be assessed at appropriate spatial scales to reflect the underlying dynamics. In this study, we systematically evaluate the performance of three popular urban mobility models, namely gravity, radiation, and visitation models across spatial scales. The results show that while the visitation model consistently performs better than its gravity and radiation counterparts, their performance does not differ much when being assessed at some appropriate spatial scale common to all of them. Interestingly, at scales where all models perform badly, the visitation model suffers the most. Furthermore, results based on the conventional admin boundary may not perform so well as compared to distance-based clustering. The cross examination of urban mobility models across spatial scales also reveals the spatial organisation of the urban structure.
\end{abstract}

\textit{Keywords:} Urban mobility model; Public transport flow; Spatial scales; Modifiable areal unit problem

\section{Introduction}

Modelling urban mobility is crucial for addressing the complex challenges cities face today, from traffic congestion and pollution to economic and social equity issues \cite{Barthelemy@2024}. Mobility shapes the accessibility of opportunities, the efficiency of urban systems, and ultimately the quality of life in cities. Well-designed transport networks enable people to reach employment, education, healthcare, and leisure activities efficiently, supporting both social and economic development. Conversely, poorly managed mobility contributes to congestion, inequities in access, and environmental stress, threatening the sustainability of cities \cite{Banister@2008,Zhao.etal@2024}. For policymakers and planners, robust mobility models are essential tools to anticipate demand, design effective infrastructure, and evaluate interventions such as transit expansion, zoning adjustments, and land-use policies.

Over the past several decades, a variety of models have been developed to explain and predict urban mobility flows~\cite{Stouffer@1940,Erlander.Stewart@1990,Simini.etal@2012,Wilson@2013,Schlapfer.etal@2021}. Among these, three approaches have gained particular prominence, namely the gravity model, the radiation model, and the more recent visitation law. Each of these models provides a distinct theoretical lens through which to understand mobility flows.

The gravity model is one of the most widely applied approaches in transportation and urban studies~\cite{Erlander.Stewart@1990}. Drawing on an analogy with Newtonian physics, it assumes that flows between two locations are proportional to their “masses” (often measured by population, employment, or activity levels) and inversely proportional to the distance between them. Owing to its simplicity and interpretability, the gravity model has been applied extensively to study commuting, migration, and trade flows~\cite{Barbosa.etal@2018}. Despite its empirical success, the model requires calibration of parameters and can oversimplify behavioural aspects of mobility.

Conversely, the radiation model was developed as an alternative to the gravity model, offering a more mechanistic, parameter-free approach to predicting human mobility \cite{Simini.etal@2012}. Instead of assuming direct distance decay, it is grounded in the concept of intervening opportunities with individuals selecting destinations based on not only their population or opportunities (e.g., jobs, amenities) but also the distribution of those within the radius between origin and destination. The model has the appeal of theoretical generality and avoids the need for calibration, though its empirical performance has been mixed across contexts. While it can capture broad mobility patterns at intercity or regional scales, its performance is often weaker in dense urban contexts where daily commuting dominates \cite{Masucci.etal@2013,Louail.etal@2015}.

The visitation law, developed more recently, draws on large-scale empirical datasets from mobile phone records and other digital traces. It describes a scaling relationship between the frequency of visits to a location and its population, capturing empirical regularities in visitation behaviour \cite{Alessandretti.etal@2020}. Unlike the gravity and radiation models, which are more mechanistic, the visitation law does not derive from physical or opportunity-based analogies but rather from observed statistical patterns. Early studies suggest that the visitation model provides a better fit to large-scale mobility data than either gravity or radiation \cite{Schlapfer.etal@2021}, though its application to public transport flows and urban-scale settings is still under active exploration.

A growing body of comparative research has examined the performance of these models in different contexts. Gravity models have often been found to outperform radiation in urban and regional contexts where commuting and daily travel dominate \cite{Masucci.etal@2013}, whereas radiation may prove more effective in interregional or migration contexts. More recent comparisons that include the visitation law report that it frequently yields the highest predictive accuracy \cite{Mepparambath.etal@2023}, capturing flow heterogeneity more effectively than either gravity or radiation \cite{Alessandretti.etal@2020,Barbosa.etal@2018}. However, findings remain context-dependent, influenced by the type of mobility data used (e.g., census commuting flows, mobile phone data, or public transport records), as well as by the spatial and temporal scales of analysis.

Despite this body of work, the role of spatial scales in shaping model performance has received comparatively little attention \cite{Yan.etal@2017}. The spatial scale at which data are aggregated and models are applied is not a neutral choice as it can fundamentally alter analytical outcomes \cite{Oshan.etal@2022,Asad.Yuan@2023,Asad.Yuan@2024}. At very fine scales, such as individual bus stops or small neighbourhoods, mobility data are often noisy, with flows subject to large fluctuations that models struggle to capture. At very coarse scales, such as large administrative regions, over-aggregation conceals meaningful local dynamics and masks heterogeneity in travel behaviour, leading to oversimplification. This issue, closely related to the modifiable areal unit problem (MAUP) in geography and spatial analysis \cite{Fotheringham.etal@1991,Goodchild@2022}, implies that the performance of mobility models is highly sensitive to the chosen scale of analysis.

Recognising the scale dependence of mobility models has two key implications. First, comparisons between models must be conducted at appropriate spatial scales to be meaningful. A model that appears to underperform at a fine scale may perform much better at an intermediate one, and vice versa \cite{Huynh@2025}. In other words, without accounting for scale, comparative assessments risk misjudging the relative strengths of models. Second, performance across scales can itself provide insight into the spatial organisation of urban systems. Dips or peaks in model performance may correspond to functional divisions of the city, such as integrated neighbourhood clusters or large-scale mobility corridors that are not immediately apparent from administrative boundaries alone. In this sense, spatial scale is not only a technical parameter but also enables models to be used as a diagnostic tool for analysing underlying urban structures.

This study addresses these gaps by systematically comparing the performance of the gravity, radiation, and visitation models across multiple spatial scales in the context of Singapore’s public transport network. The role of spatial scales in examined by employing spatial units constructed using two different methods. First, we use the conventional administrative boundaries defined by the Urban Redevelopment Authority~\cite{URA@2019}, which divide Singapore into regions, planning areas, and subzones for the purposes of urban governance and planning. Second, we construct distance-based clusters of transport nodes (bus stops and train stations) using Voronoi tessellation and distance-based clustering \cite{Huynh@2025}. This data-driven method provides an alternative representation of space that more closely reflects the actual organisation of movement patterns in the city. By comparing these two approaches, we assess not only the models themselves but also the implications of how spatial units are defined.

The contributions of this study are threefold. First, we provide a comprehensive comparison of gravity, radiation, and visitation models using public transport flow data, extending previous work that often focused on a single model or limited datasets. Second, we explicitly examine the role of spatial scale, offering empirical evidence of its impact on model performance and highlighting the need for scale-sensitive evaluation. Third, we contrast conventional administrative boundaries with distance-based spatial units, showing conventional boundaries may obscure important mobility structures and that data-driven approaches can more accurately capture mobility dynamics. Together, these findings advance methodological understanding of mobility modelling and carry practical relevance for urban and transport planning.

In the remainder of the paper, we first describe in Sec. \ref{data_method} the data and method used for the analysis, including details on public transport flows, the construction of spatial units, and the implementation of the three models. After that, we report in Sec. \ref{results} the results of the analysis and discuss their implications, as well as the strengths and limitations of the current study. Finally, we conclude with key findings, practical insights, and potential directions for future studies on this topic.

\section{Data and method}
\label{data_method}
\subsection{Data}

This study uses the public transport trips data from Singapore to demonstrate the performance of the urban mobility models at different spatial scales. Singapore's public transportation system is a highly integrated, multi-modal network designed to provide efficient and accessible mobility for its urban population. The system is a cornerstone of the city-state's land use and transport planning strategy, aiming to reduce reliance on private vehicles and manage traffic congestion in the land scarce city-state. The Singapore public transport system comprises of the Mass Rapid Transit (MRT), Light Rail Transit (LRT) and the public bus networks. The public transport trips data from Singapore is available in the transport node level, whereby it provides the number of trips between different bus stop or train station pairs within the bus and train network in Singapore \cite{DataMall}. 

\subsection{Spatial scales}

Modelling urban mobility requires first defining the spatial units that represent places, since these units serve as the basis for calculating proxies of activities and opportunities. A common and convenient approach is to adopt the administrative boundaries established by planning authorities. In Singapore, the Urban Redevelopment Authority (URA) employs a hierarchical zoning system to guide development and infrastructure provision \cite{URA@2019}. At the broadest level, the island is divided into five planning regions, which are further subdivided into 55 planning areas serving as the primary divisions used in urban planning and census reporting. Each planning area is envisioned as a self-contained town with a balanced mix of residential, commercial, and recreational functions. At a more granular scale, planning areas are broken down into subzones, which often revolve around neighbourhood centres or local activity nodes, enabling fine-grained planning policies.

While administratively convenient, these boundaries may not fully capture the dynamics of urban mobility as people do not necessarily travel according to imposed planning zones, but instead move along patterns shaped by the transport network. To better reflect these dynamics, our study also employs data-driven, distance-based clustering of transport nodes as an alternative to administrative divisions. The process begins by dividing Singapore into Voronoi polygons, each associated with a single transport node (bus stop or train station). Voronoi tessellation ensures that every location in the city is assigned to its nearest node, producing a partition of space that reflects accessibility to the public transport network. Building on this foundation, transport nodes are then clustered using the procedure described in \cite{Huynh@2025}. Specifically, a distance threshold is introduced to determine whether two nodes should belong to the same cluster. Smaller thresholds produce many small clusters, capturing fine-grained local interactions, while larger thresholds yield fewer, larger clusters that aggregate broader patterns. In this way, the threshold distance serves as a control parameter for spatial scale. Once clusters are formed, the Voronoi polygons of the constituent nodes are merged to generate spatial units representing each cluster.

Using these two approaches of administrative boundaries and distance-based clustering, we systematically assess the impact of spatial scale on the performance of three urban mobility models, namely the gravity law, radiation law, and visitation law. This approach allows us to compare imposed and organic representations of space, thereby evaluating not only the models themselves but also the ways in which spatial definitions shape our understanding of urban mobility patterns.

\begin{figure}
\centering
\includegraphics[width=\textwidth]{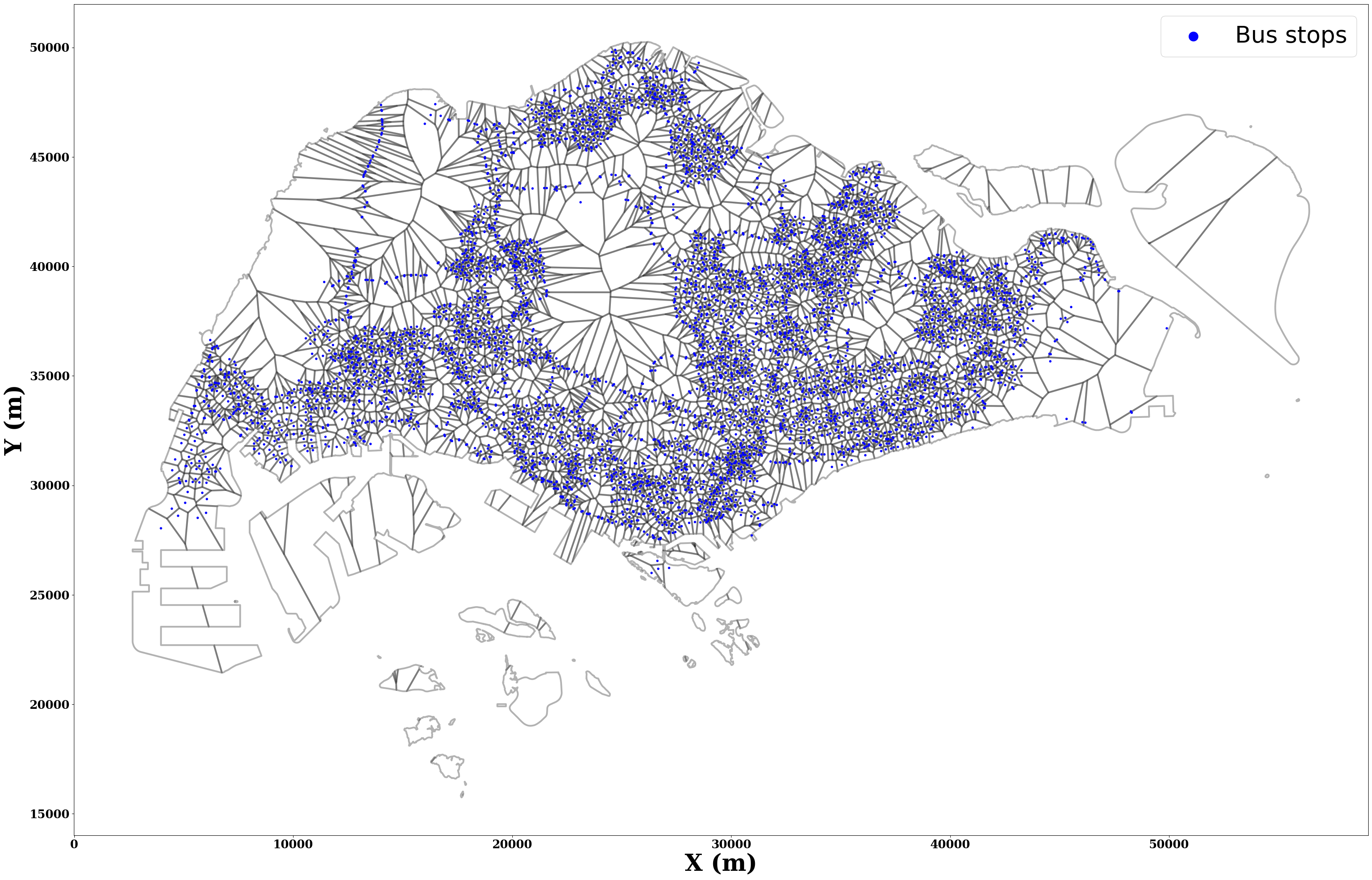}
\includegraphics[width=\textwidth]{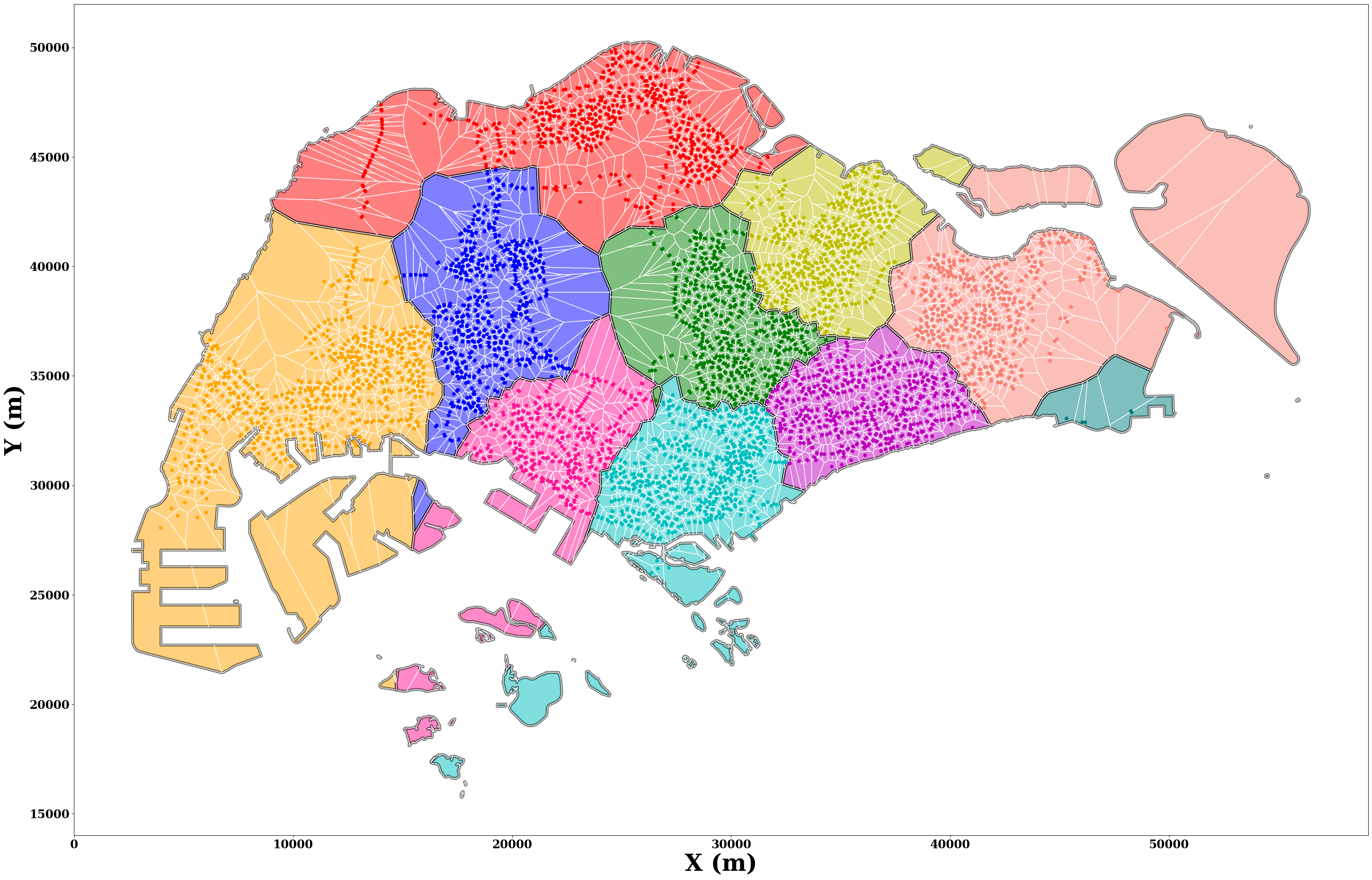}
\caption{Construction of the spatial units at different spatial scales. Left: Voronoi tessellation of individual transport nodes (bus stops and train stations). Right: clusters of nodes (colour coded, following the procedure in \cite{Huynh@2025}) at 3000 m together with the merging of corresponding Voronoi cells.\label{voronoi_cells}}
\end{figure}

\subsection{Urban mobility models}
This section provides the details of the three models discussed in the study.

\subsubsection{Gravity model}
The gravity model is a common urban mobility model that predicts the number of trips between two urban zones based on their ``attractiveness'' and the ``friction'' of the distance or travel time between them. It is a fundamental component of the traditional four-step transportation planning model.\par

The model is named after Newton's law of universal gravitation, as it assumes that the interaction (or trip volume) between two locations is directly proportional to their ``masses'' (e.g., population, employment) and inversely proportional to the square of the distance between them. In urban mobility, the ``mass'' of a zone is its ability to produce and attract trips, while the ``distance'' is a generalised measure of travel impedance. A simpler form of gravity model, where distance is used as the travel impedance can be expressed as
\begin{equation}
    \label{Eqn_GravityModel}
    T_{ij} = k \frac{P_i^{\alpha} A_j^{\beta}} {r_{ij}^{\gamma}} \,,
\end{equation}
where $P_i$ is the variable that influence the trip production from origin zone $i$, $A_j$ is the variable that influence the trip attraction to destination zone $j$, $r_{ij}$ is the distance between origin $i$ and destination $j$ and $k, \alpha, \beta, \gamma $ are the parameters calibrated using data.
While intuitive and widely used for its simplicity, the gravity model has limitations. The model describes observed patterns but doesn't explain the underlying behavioural reasons for travel. It can also oversimplify complex human behaviours and may not perform well at fine spatial scales, where factors like intervening opportunities or local amenities become more significant than simple distance and population.
\subsubsection{Radiation model}
The radiation model is a parameter-free model for urban mobility that predicts travel flows between locations based on the principle of intervening opportunities. Unlike the gravity model, which relies on a distance decay parameter that needs to be calibrated, the radiation model requires only the population distribution of a region to predict human movement. The core idea is that an individual seeking a destination (like a job) will consider all opportunities within a certain distance, but their choice is influenced by the number of opportunities that are closer to them. The model, therefore, captures the idea that a person is less likely to travel to a distant location if there are a sufficient number of good opportunities available closer to their home. Mathematically, the radiation model is expressed as
\begin{equation}
    \label{Eqn_RadiationModel}
    T_{ij} = T_i \frac{m_in_i}{(m_i+s_{ij})(m_i+n_j+s_{ij})} \,,
\end{equation}
where $T_i$ is the predicted number of trips from an origin zone $i$ to a destination zone $j$, $T_i$ is the total number of trips originating from zone $i$, $m_i$ and $m_j$ are the populations (or a measure of opportunities, such as employment) in zone $i$ and zone $j$, respectively, $s_{ij}$ is the total population of all zones located within a circle centered at zone $i$ with a radius equal to the distance between $i$ and $j$, excluding the populations of zones $i$ and $j$. This represents the ``intervening opportunities'' that could ``absorb'' the trips.\par

A key advantage of the radiation model is its parameter-free nature, which makes it a more universal and robust tool for predicting mobility in different regions or at different scales without extensive data calibration. Studies have shown that it often provides more accurate predictions for long-distance commuting and migration patterns than the gravity model. However, it may be less effective for predicting short-distance, intra-urban trips where factors beyond just population and intervening opportunities, like specific land use or local amenities, play a larger role.

As we need to identify the amount of opportunities $m_i$, $n_j$ at different levels of spatial aggregation, we employ the Voronoi tessellation to divide the land surface. The Voronoi tessellation is defined based on the clusters of nodes at every level of spatial aggregation without leaving gaps so that all opportunities are taken into account and assigned to the corresponding cluster (see Fig. \ref{voronoi_cells}).

\subsubsection{Visitation model}
The universal visitation law is a recently discovered scaling law in urban mobility that describes a fundamental relationship between an individual's travel distance and their visiting frequency to any urban location. This law provides a simple, robust framework for understanding and predicting the patterns of human movement in cities. The law states that the number of visitors to a location is governed by a simple inverse square relationship with the product of their visiting frequency and travel distance. In simpler terms, a location can attract the same number of people from far away who visit infrequently as it does from nearby people who visit often. \par
This empirical law is theoretically explained by the Exploration and Preferential Return (EPR) model of human mobility. This model suggests that human movement is a two-part process: individuals first explore new, previously unvisited locations and then preferentially return to a small number of frequently visited places. The universal visitation law emerges from this interplay, as the trade-off between exploring distant, novel locations and returning to familiar, close-by ones balances out in a predictable way.\par
The visitation law is given by the following expression
\begin{equation}
    \label{Eqn_VisitationLaw}
    V_{i}(r, f) = \frac{\mu_i}{(rf)^\eta} \,,
\end{equation}
where $V_{i}(r, f)$ is the number of visitors to a location $i$, $r$ is the travel distance from a person's home to the location, and $f$ is the frequency of visits to that location. $\eta$ is the scaling exponent, which has been found to be approximately 2 across various cities, and $\mu_i$ is the proportionality constant that adjusts the model to reflect the specific attractiveness of the location.
Building on some approximations on the universal visitation law, Schläpfer et al. (2021) also introduced a method to calculate aggregate origin-destination (O-D) flows using population density. This is expressed by the following equation.

\begin{equation}
    \label{Eqn_VisitationLaw_ODflow}
    T_{ij} = \frac{\mu_j A_i + \mu_i A_j}{r_{ij}^2 \ln\left(\frac{f_{max}}{f_{min}}\right)} \,,
\end{equation}

The proportionality constants $\mu_i$ and $\mu_j$ capture the location-specific attractiveness of the origin $i$ and destination $j$. Specifically, $\mu_j$ is expressed as
\begin{equation}
\mu_j \approx \rho_{pop} (j) r_j^2 f_{home}
\end{equation}
where $\rho_{pop} (j)$ is the population density at the destination $j, r_j$ is the distance to the boundary of  $j$ and $f_{{home}}$ is the frequency of returning to home during the observation period. A similar formulation can be used to derive $\mu_i$. The home-return frequency is approximated as $f_{home}\approx1/day$, under the assumption that individuals return home on a daily basis. The areas of location $i$ and $j$ are denoted by $A_i$ and $A_j$, respectively. The visitation frequency is constrained by two limits: the minimum frequency $f_{min}=1/T$ where $T$ is the observation period and the maximum frequency $f_{max}=1/day$. Further details on the visitation law for aggregate flows can be found in \cite{Schlapfer.etal@2021}.

The discovery of this law, based on large-scale mobile phone data, provides a new, universal metric for urban dynamics. Unlike other mobility models that focus on aggregate flows between zones, the visitation law focuses on the behavior of individual visitors. It has significant implications for urban planning, commercial site selection, and understanding the resilience of cities to disruptions, as it offers a fundamental, data-driven way to quantify and predict how people interact with the urban landscape.

\subsection{Evaluation of models' performance}

In order to evaluate the performance of the models, we employ the adjusted $R^2$ as a performance indicator to account for the sample size and complexity of the model. It is calculated based on the conventional coefficient of determination $R^2$
\begin{equation}
\label{adjusted_R}
R_{adj}^2=1-\frac{n-1}{n-p-1}(1-R^2)
\end{equation}
which approximates $R^2$ if $n\gg p$, with n being the sample size and p the number of parameters in the model.  It should, however, be noted that the adjusted $R^2$ only differs in the case of the gravity model because the radiation and visitation models are parameter-free, making $R^2$ identical to $R_{adj}^2$ with $p=0$ in Eq. \ref{adjusted_R}.

The model is trained on 50\% of the data and tested on the remaining 50\% based on which $R^2$ is calculated. For each configuration of the spatial units, the model is run 100 times with randomisation of the training and testing portion of the data. The adjusted $R^2$ is calculated for each run and its mean value and standard deviation is computed over the 100 runs.

\section{Results and discussion}
\label{results}
\subsection{Performance of the models across spatial scales}

Our comparative analysis of the gravity, radiation, and visitation models across spatial scales reveals distinct differences in their ability to capture urban mobility patterns. Overall, the visitation law consistently achieves the best performance, followed by the gravity model and, lastly, the radiation model. This ordering aligns with earlier findings in the literature \cite{Mepparambath.etal@2023}, reinforcing the robustness of the visitation law in explaining aggregate travel flows.

\begin{figure}[t]\vspace*{4pt}
\includegraphics[width=\textwidth]{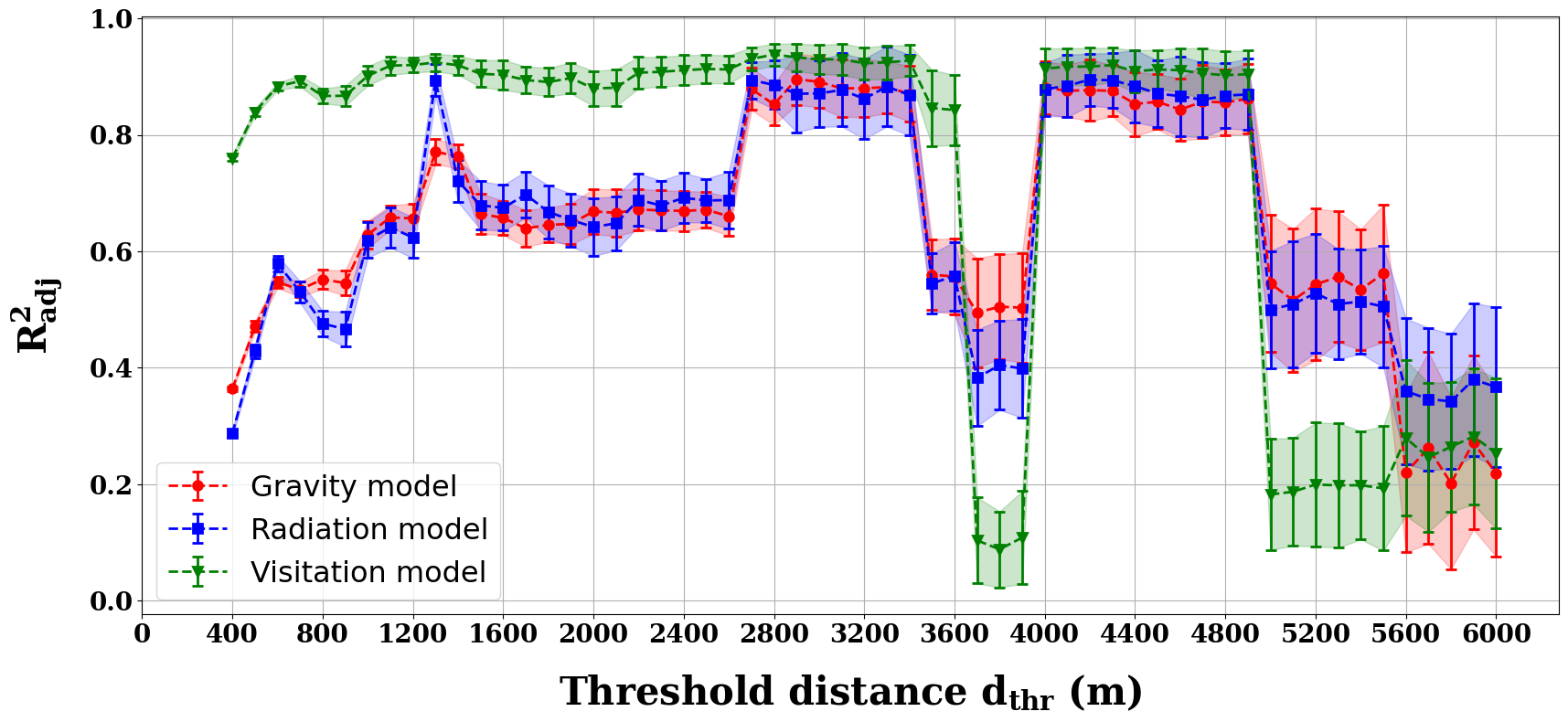}
\caption{Performance of the urban mobility models across different spatial scales.\label{results_distance}}
\end{figure}

At fine spatial scales, both the gravity and radiation models perform poorly. The abundance of origin–destination (OD) pairs at this resolution creates noisy dynamics, as each pair covers relatively small areas and populations that are highly susceptible to fluctuations. By contrast, the visitation model shows greater resilience under such conditions, maintaining relatively strong predictive power. As spatial scale increases, all three models exhibit improvement in their performance, suggesting that moderate aggregation helps mitigate noise while preserving meaningful mobility patterns (see Fig. \ref{results_distance}).

Across most spatial scales, the visitation model outperforms the other two, although at their respective optimal scales the three models converge to similar adjusted $R^2$ values, indicating comparable explanatory power. Notably, all three models reach peak performance at an intermediate scale of approximately 3,000 m, which appears to strike a good balance between granularity and generalisation. This spatial resolution allows the models to best capture the underlying dynamics of mobility flows without overfitting or excessive loss of detail.

At larger spatial scales, performance declines for all models due to over-aggregation, which obscures variability in travel behaviour between different locations. Additionally, there are specific spatial windows in which all models perform poorly. Interestingly, within these windows, the visitation law performs worse than the gravity and radiation models. This suggests that the relative strengths of different modelling approaches may vary depending on the spatial context, highlighting the importance of scale sensitivity in model evaluation. These findings suggest that urban mobility models inherently operate at different effective spatial scales. Recognising this scale dependence is crucial when assessing their validity and practical application. The results also suggest that selecting an appropriate level of spatial aggregation is as important as the choice of model itself in ensuring accurate and meaningful interpretations of urban travel behaviour.

\subsection{Comparison of administrative boundary and distance-based clustering}

\begin{figure}[t]\vspace*{4pt}
\includegraphics[width=\textwidth]{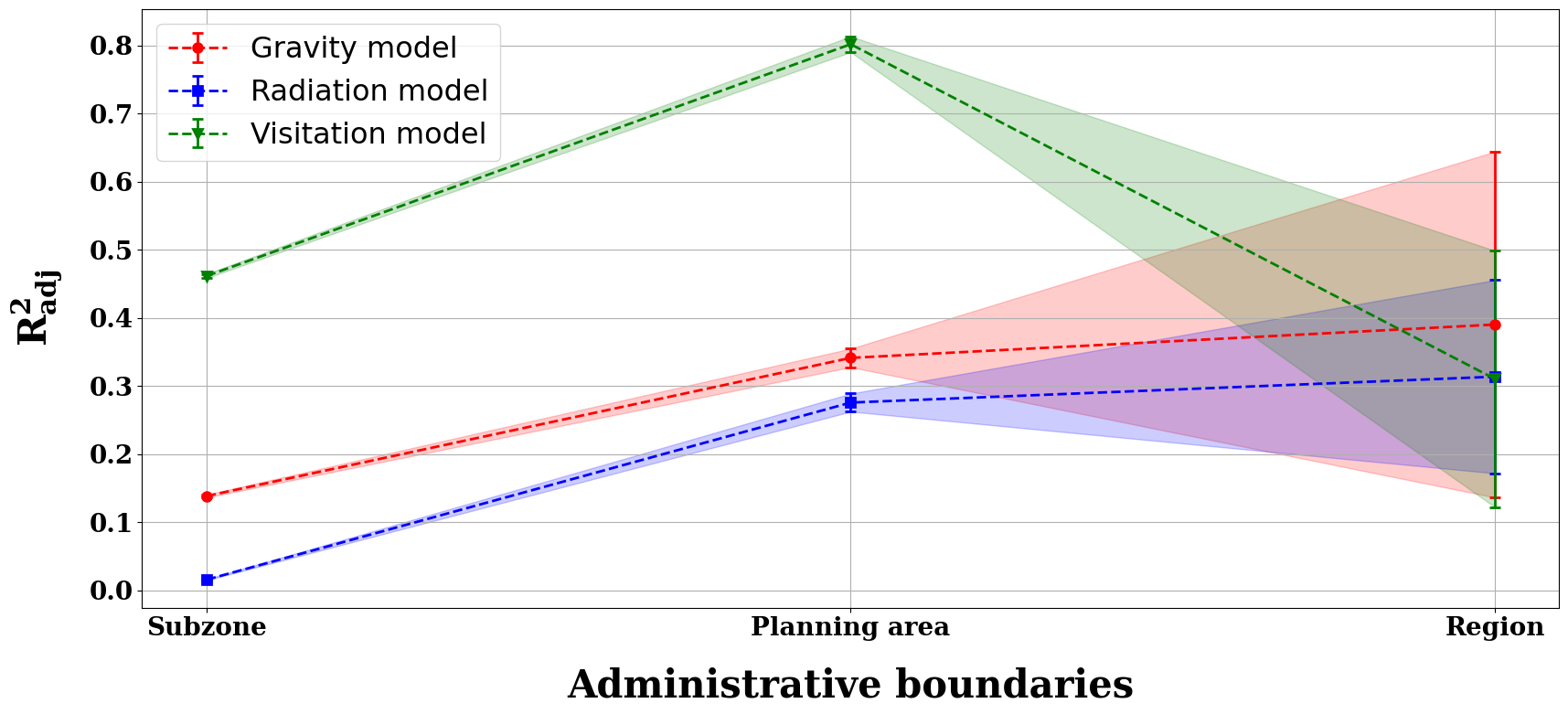}
\caption{Performance of the urban mobility models at different levels of administrative boundary.\label{fig_admin_boundary_results}}
\end{figure}

For comparison with distance-based spatial aggregation, the performance of the three mobility models using different administrative boundary levels in Singapore (subzone, planning area, and region) is shown in Fig. \ref{fig_admin_boundary_results}. Across these boundaries, the models exhibit a consistent pattern with poorer performance at both finest and coarse scales, and improved performance at some intermediate scale. Among the three levels of administrative boundary, the planning area generally yields higher explanatory power compared to subzone and region, suggesting its offering of a more balanced resolution. This pattern likely arises because subzones are too granular, resulting in noisy dynamics dominated by small populations and volatile trip counts, whereas regions are too coarse, leading to over-aggregation and the masking of local variabilities. Planning areas thus represent a compromise between detail and generalisation. However, this should not be taken as evidence that planning areas represent an optimal scale; rather, subzones may be too fine and regions too coarse, with planning areas falling somewhere in between.

When compared directly with distance-based clustering, however, the limitations of administrative units become clear. At their best-performing administrative level (planning areas), all models still underperform relative to distance-based aggregation. For a more direct comparison, we consider the spatial scale that most closely corresponds to planning areas, around 600 m \cite{Huynh@2025}. The models at this scale, although not at their best, still perform better than they do under the planning area boundary. Similarly, at the scale corresponding to regions (around 4,400 m \cite{Huynh@2025}), the models also perform substantially better than when using the region boundary. These results indicate that imposed administrative units do not fully reflect actual mobility patterns and are therefore not necessarily the most suitable for evaluating urban mobility models. Conversely, clustering transport nodes by spatial proximity appears to offer a more faithful representation of mobility patterns than conventional administrative divisions.

This discrepancy has important implications. Because model performance depends on the spatial framework used, evaluations based solely on administrative boundaries, while convenient, may understate the explanatory power of mobility models or distort comparisons between them. These boundaries are imposed constructs that may not align with organic patterns of movement within the city. Relying on them without scrutiny risks underestimating the performance of mobility models or misinterpreting their comparative effectiveness. By contrast, distance-based clustering adapts more flexibly to the natural geography of mobility, yielding results that better capture the true interaction patterns among locations. This underscores that both the choice of model and the choice of spatial units are critical in shaping the quality and interpretation of urban mobility analysis.

\subsection{Implications for urban planning}

The results of this study could have important implications for urban and transport planning, particularly in understanding how people actually move across the city. The mobility patterns revealed through the models highlight the existence of functionally integrated zones that extend beyond imposed administrative boundaries. For example, areas such as Jurong East and Bukit Batok in the west, or Woodlands, Yishun, and Sembawang in the north, emerge as coherent clusters of mobility activity. These areas are not simply administrative units but functional regions shaped by travel flows, suggesting that urban residents perceive and use the city in ways that are not neatly captured by statutory borders. Recognising such integrated zones can help planners identify where resources, services, and infrastructure might be shared or coordinated more effectively.

An interesting observation in our analysis is the dip in model performance around the 3,700–3,900 m scale. At this spatial threshold, the explanatory power of all models decreases noticeably, despite the general trend of improvement with increasing aggregation before the 3,500 m scale. This suggests that at around 4 km, clusters of transport nodes may cross into multiple functional zones, aggregating flows that do not belong to a single coherent mobility system. As a result, the models struggle to capture consistent travel patterns because the aggregation mixes movements that are functionally distinct. This dip illustrates that there are critical thresholds where spatial aggregation can inadvertently blur the natural organisation of mobility, leading to misrepresentation of flows. For planners, this means that scale effects cannot be overlooked when choosing the wrong level of aggregation can obscure rather than reveal meaningful patterns of urban activity.

These insights can directly inform transport and land-use planning. By identifying natural functional areas defined by mobility flows, planners can design policies and infrastructure that better align with actual travel behaviour. For example, improving public transport connectivity within integrated clusters may have greater impact than reinforcing links that merely follow administrative borders. Similarly, land-use planning can benefit from recognising mobility-defined zones, ensuring that housing, employment centres, and amenities are distributed in ways that correspond to how people already move through the city.

Ultimately, the findings demonstrate that mobility-driven functional regions provide a more accurate basis for planning than administrative divisions. By incorporating these insights into planning strategies, cities can better allocate resources, improve accessibility, and foster more efficient and inclusive urban systems that reflect the lived realities of residents.

\subsection{Strengths and limitations}

A key strength of this study lies in its systematic evaluation of multiple urban mobility models across spatial scales using detailed public transport flow data. By comparing gravity, radiation, and visitation models, the analysis provides a nuanced understanding of how model performance varies not only with scale but also with the choice of aggregation method. This comparative framework allows us to highlight the relative advantages of each modelling approach and the conditions under which they are most effective, offering practical insights for both researchers and planners.

Nevertheless, several limitations should be acknowledged. First, the reliance on population data as a proxy for activity levels may not accurately represent the true intensity of mobility-generating locations. Areas such as Singapore’s central business district (CBD) have relatively low residential populations but generate disproportionately high traffic flows due to employment, commercial, and leisure activities. Using population as the sole indicator therefore risks underestimating flows in such zones, leading to model biases. Incorporating alternative measures of activity, such as employment density or land use mix could address this limitation.

Second, the study employs Euclidean distance as the measure of separation between locations. While this offers computational simplicity and consistency, it does not reflect the reality of urban transport networks. Travel decisions are shaped more directly by network distance (actual paths along roads and rail) or, more importantly, travel time, which accounts for congestion, service frequency, and connectivity. As a result, Euclidean distance may misrepresent accessibility in cases where physical barriers or indirect routes significantly alter effective travel costs. Future research could strengthen model accuracy by incorporating network-based or time-based measures of distance, which would more closely align with how commuters perceive and experience mobility.

\section{Conclusion}
This study systematically assessed the performance of three widely used urban mobility models, namely the gravity, radiation, and visitation models, across a range of spatial scales. By aggregating population according to clusters of transport nodes at varying distance thresholds, we revealed how scale fundamentally shapes model performance. The results show that the visitation model consistently outperforms the gravity and radiation models at most scales. Yet, at their respective optimal scales, the differences between models narrow considerably, with all three achieving similar levels of explanatory power. Interestingly, there exists a window of spatial scales where all models perform poorly, and within this window the visitation law suffers most, suggesting the presence of a deeper urban organisational structure that constrains mobility dynamics in Singapore.

A key finding is that the effectiveness of mobility models is highly scale-dependent. Selecting inappropriate spatial scales can distort model evaluation and obscure their comparative strengths. An optimal intermediate scale exists at which the models best capture urban dynamics, balancing detail against generalisation. Moreover, the analysis shows that performance based on administrative boundaries is generally inferior to that based on distance-based clustering. While convenient, imposed boundaries do not necessarily align with actual travel behaviours, potentially masking critical patterns of interaction. By contrast, distance-based aggregation more accurately reflects the functional structure of urban mobility, providing a clearer lens on how the city operates.

The performance of mobility models across scales not only highlights the importance of scale-aware analysis but also reveals the organisational structures embedded within urban systems that dictate large-scale travel flows. For transport and land-use planning, this underscores the need for data-driven approaches that recognise functional regions beyond administrative borders. Future research could extend this work by incorporating additional determinants of mobility, such as land-use patterns, employment distribution, or the structural properties of the transport network. Integrating such factors would further enhance the explanatory power of urban mobility models and provide deeper insights for designing inclusive, efficient, and resilient urban transport systems. Ultimately, understanding and accounting for scale effects is as important as the choice of model itself when analysing and planning urban mobility.

\section*{Acknowledgement}
This research is supported by A*STAR project number CoT-H1-2025-3 under the Cities of Tomorrow Grant 2024.

\bibliography{ref}
\bibliographystyle{unsrt}

\end{document}